\documentclass[prb,showpacs,a4paper,twocolumn,superscriptaddress]{revtex4}
\usepackage{amsfonts,amssymb,amsmath}
\usepackage[english]{babel}
\usepackage{graphicx}
\usepackage{latexsym}
\usepackage{setspace}
\usepackage{pstricks}

\begin{document}


\author{N. Spallanzani}
\affiliation{Department of Physics, University of Modena e Reggio Emilia, via Campi 213a, 41100 Modena, Italy}
\affiliation{INFM-CNR National Research Center "S3", Modena, Italy}
\email{nicola.spallanzani@unimore.it}

\author{C. A. Rozzi}
\affiliation{INFM-CNR National Research Center "S3", Modena, Italy}
\email{rozzi@unimore.it}

\author{D. Varsano}
\affiliation{INFM-CNR National Research Center "S3", Modena, Italy}
\affiliation{European Theoretical Spectroscopy Facility (ETSF)}

\author{T. Baruah}
\affiliation{Department of Physics, University of Texas at El Paso, El Paso, Texas 79968}

\author{M. R. Pederson}
\affiliation{Center for Computational Materials Science, Code 6390, Naval Research Laboratory, Washington, DC 20375-5345}

\author{F. Manghi}
\affiliation{Department of Physics, University of Modena e Reggio Emilia, via Campi 213a, 41100 Modena, Italy}
\affiliation{INFM-CNR National Research Center "S3", Modena, Italy}

\author{A. Rubio}
\affiliation{Nano Bio Spectroscopy group, Dpto. de F\'{\i}sica de Materiales, Universidad del Pa\'{\i}s Vasco and Centro Mixto CSIC-UPV/EHU and DIPC, Av. Tolosa 72, E-20018 San Sebasti\'an, Spain}
\affiliation{European Theoretical Spectroscopy Facility (ETSF)}


\title{Photo-excitation of a light-harvesting supra-molecular triad: a Time-Dependent DFT study}

%

%
%


\begin{abstract}

We present the first time-dependent density-functional theory (TDDFT)
calculation on a light harvesting triad carotenoid-diaryl-porphyrin-C$_{60}$.
Besides the numerical challenge that the {\it ab initio} study of the electronic
structure of such a large system presents, we show that TDDFT is able to provide
an accurate description of the excited state properties of the system. In
particular we calculate the photo-absorption spectrum of the supra-molecular
assembly, and we provide an interpretation of the photo-excitation mechanism in
terms of the properties of the component moieties. The spectrum is in good
agreement with experimental data, and provides useful insight on the
photo-induced charge transfer mechanism which characterizes the system.

\end{abstract}

\maketitle

\section{Introduction}

The quest for efficient ways to exploit renewable energy sources has become one
of the key problems of our times. The idea of a solid-state device capable of
transforming solar light energy into electric potential energy is quite old, and
enormous progress has been made in optimizing traditional junction-based devices
with respect to their thermodynamical efficiency and production cost
\cite{green04}. Recently entirely new classes of materials have been proposed as
building blocks for next-generation solar cells. Among these nanostructured
semiconductors, organic-inorganic hybrid and supra-molecular assemblies appear
to be the most promising \cite{energy_review}.

In the class of supra-molecular assemblies the donor-acceptor dyads and triads
are objects of careful experimental study. They are in fact considered an ideal
mimic for the primary photo-synthetic process, which basically consists of a
photo-excitation followed by a charge-transfer between the component units
\cite{photosynthesis}. The same process on the other hand can be exploited in
photo-conversion units like dye-sensitized (or Gr\"atzel) solar cells
\cite{gratzel91,gratzel01}. The research for ideal components of
light-harvesting dyads has led to porphyrin-C$_{60}$ assemblies \cite{antenna}.

Porphyrin molecules show extensive absorption of visible light, and an optical
gap of approximately 2 {\sl eV} and are therefore ideal photo-reaction
centers and electron donors. C$_{60}$, on its side, is the ideal electron
acceptor. It has a particularly deep triply degenerate LUMO at approximately
-4.3 {\sl eV} vs vacuum, capable of accommodating six electrons from neighboring
molecules. Moreover the highly conjugated nature of its bonds, together with a
cage radius of 3.45 \AA, stabilizes the extra charges in a very efficient way,
so that little repulsive forces are felt by the excess electrons. A further step
toward solar cells based on these components consisted in transforming the
donor-acceptor unit into a donor-bridge-acceptor unit. This change has the main
purpose of obtaining a better charge separation and a longer life time of the
final (excited) state.

While many experimental properties of porphyrin-C$_{60}$ based dyads and triads
are known, less is known from an {\em ab-initio} theoretical point of view. The
main reason is that the large size of the systems yields traditional accurate  
quantum-chemistry methods such CI impractical or infeasible. Less accurate but
computationally efficient methods such as DFT have successfully been employed to
describe the ground state of such compounds \cite{baruah_triad}, but
unfortunately, the prediction of excited state properties, like the optical
absorption spectrum, from the ground state Kohn-Sham orbitals and eigenvalues
usually leads to results that are in discord with the observed spectra. In
particular the role of the interfaces between different parts of the triad and
between the molecule and the solvent are not understood in spite of the fact
that they are of paramount relevance to the problem of improving the energy
conversion efficiency of next generation solar cells.

TDDFT is a rigorous and in principle exact method that allows for the
calculation of excited-states through an extension of DFT to the domain of
time-dependent external potentials \cite{tddft}. In this letter we report the
first TDDFT calculation of the electronic structure and absorption spectrum of
Carotenoid-diaryl-Porphyrin-C$_{60}$ triad (abbreviated hereafter as {\sl triad}
or C-P-C$_{60}$). 

\section{Results and Discussion}

\begin{figure}[h]
\includegraphics[width=0.9\textwidth]{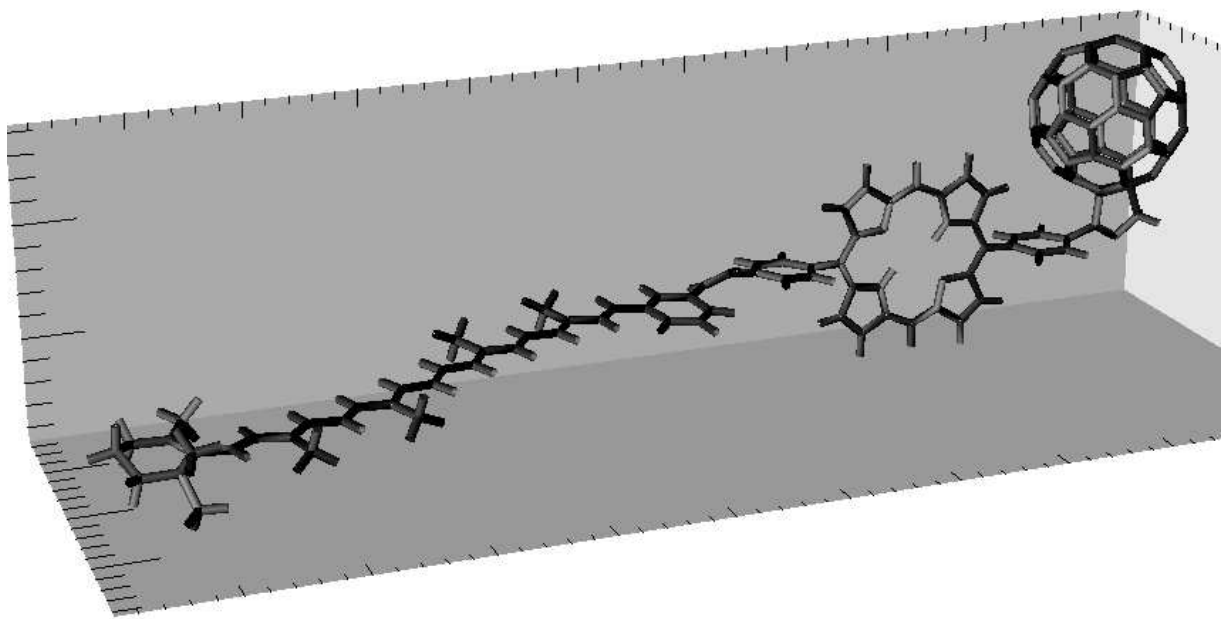}
\caption{Optimized geometry of C-P-C$_{60}$ as in Ref.~\citep{baruah_triad}. Minor tics are 1 \AA\ wide apart.}\label{fig:geometry}
\end{figure}

The C-P-C$_{60}$ system is composed by 207 atoms with 632 valence electrons. Its
structure corresponds to the formula C$_{132}$H$_{68}$N$_6$O. For all our
calculations we used an optimized geometry obtained by Baruah {\em et al.} using
the LBFGS scheme \cite{baruah_triad}. The earlier calculation by Baruah and
Pederson was carried out at the all-electron generalized gradient level using 
the NRLMOL code. This code uses a large gaussian basis set within the LCAO 
formulation. In this structure the molecule can be entirely embedded in a box
approximately $55\times 16\times20$ \AA\ wide, the longest side being along the
{\sl x} axis. As it is also clear following a steric argument, the porphyrin
ring is perpendicular to the aryl rings, which, in turn, are coplanar with the
carotenoid. The optimized geometry shows a slight bend in the carotenoid chain
(See~\ref{fig:geometry}).

Our first goal was to study the ground state properties of the molecule within
the same methodology that we use for the time-dependent calculation of the
excited states, and compare the electronic structure of the triad to the
electronic structures of the separated components. For the calculations of the
separated moieties the same geometries used in the triad have been utilized.

We have first performed a standard DFT calculation at the LDA level, using the
Perdew-Zunger correlation functional \cite{perdew-zunger81}. The core electrons
were represented via Troullier-Martins\cite{troullier-martins91} 
pseudo-potentials \footnote{For all the components the pseudo-potentials cutoff radii are  1.25 a.u. for hydrogen atoms, 1.47 a.u. for carbon atoms, and 1.39 a.u. for nitrogen and oxygen atoms.}, and the Interpolating Scaling
Functions method \cite{isf}  was used to efficiently solve the Poisson's
equation in the calculation of the Hartree potential. All the simulations in the
present work, apart from the geometry optimization, were performed discretizing
all the quantities in real-space over a uniform mesh made of overlapping
spheres with a radius of 4.0 \AA\ centered around each nucleus. The adopted grid
spacing of 0.2 \AA\ guarantees the convergence of the total energy of the system
\cite{oct06}.

\begin{figure}[h]
\includegraphics[width=0.9\textwidth]{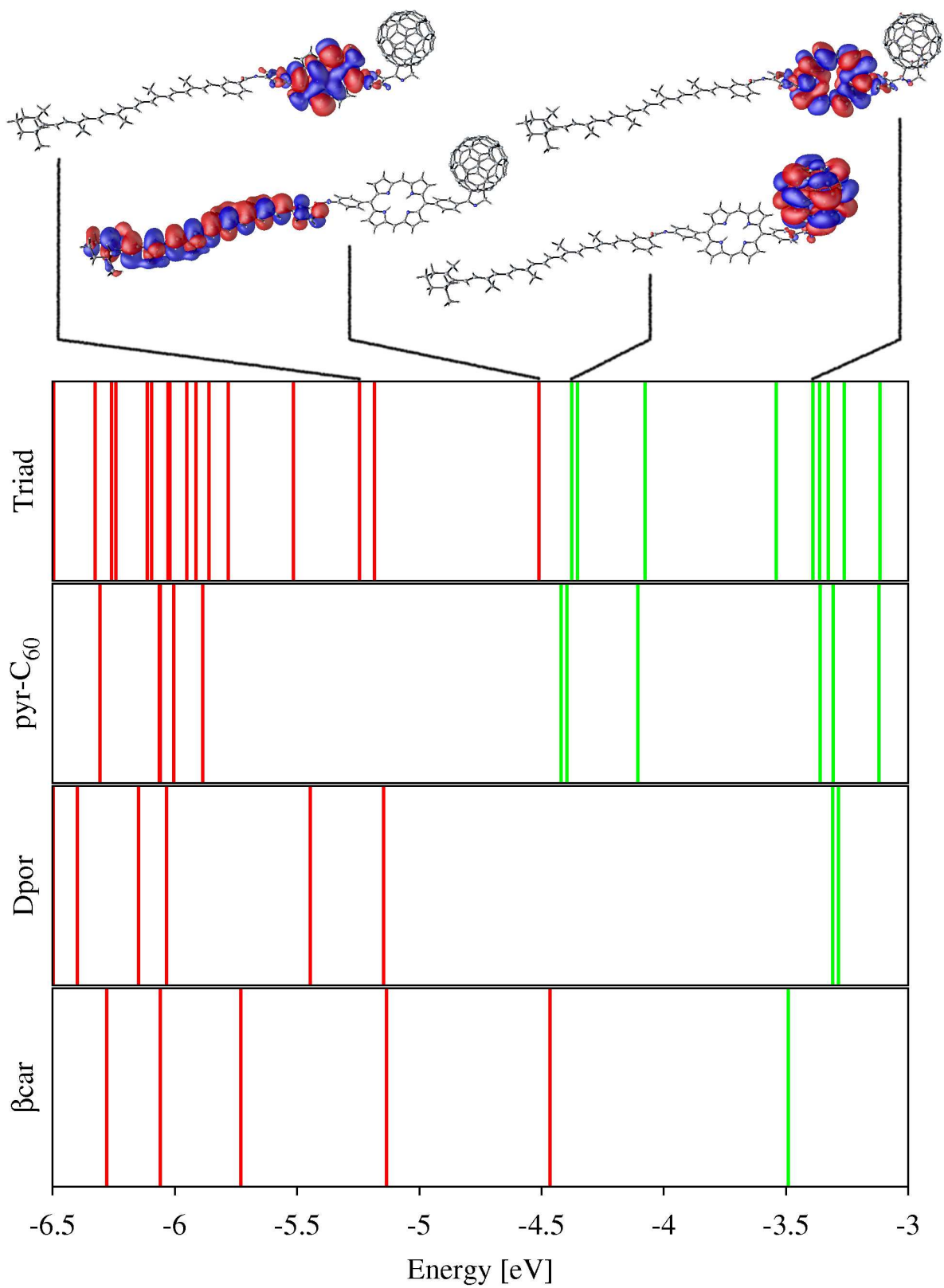}
\caption{Ground state energy levels of the triad. Red occupied, green unoccupied states. The orbitals involved in the photo-absorption and charge transfer processes are shown on top.}\label{fig:triad_levels}
\end{figure}

In \ref{tab:en_lev} we report the Kohn-Sham energies of HOMO and LUMO
for the isolated moieties and for the full triad. The detailed energy levels
for other orbitals are depicted in~\ref{fig:triad_levels}.

Comparing the values of the HOMO and LUMO energies of the triad with those of
the isolated moieties we notice that the energy of the HOMO in the triad and in
the isolated $\beta$-carotenoid differ by only 40 {\sl meV}, as well as the
energy of the LUMO in the triad and in the isolated pyrrole-C$_{60}$. It is also
possible to find a close correspondence between many of the orbitals localized
on each molecule, and orbitals of the full triad. In particular the HOMO
wave-function of the triad is localized on the carotenoid, and it corresponds to
the HOMO of the isolated $\beta$-carotenoid, while the LUMO of the triad is
fully localized on the C$_{60}$ and it corresponds to the LUMO of the C$_{60}$
alone (see~\ref{fig:triad_levels}). These features are in agreement with
the earlier results obtained in Ref.~\citep{baruah_triad} using an all-electron GGA approach.

In some other cases the correspondence between the localization of orbitals
remains, but some degeneracies are lifted, for example the triply degenerate
C$_{60}$ LUMO is split into three separate levels in the triad. Finally some
orbitals present a mixed character, such as the LUMO+5, LUMO+6, LUMO+7 with
mixed P-C$_{60}$ character, or HOMO-6, with mixed C-P character. These orbitals
however do not appear to have a major role in the photo-excitation process (see
below).

From the above results we conclude that the electronic structure of the ground
state of the triad is well approximated by the sum of the electronic structures
of the isolate molecules. This conclusion implies that it is possible to design
triad-based photoconversion devices adopting a {\sl divide et impera} strategy,
by covalently joining pre-optimized molecular components.


We have calculated the optical absorption spectra of the triad and its
separated components using the time-dependent DFT as described in
Ref.~\citep{tddftbook}. 
Starting from the ground state Kohn-Sham orbitals $\tilde{\phi}_j$, the system is
instantaneously perturbed with a weak external electric potential 
of magnitude $k_0$ along the principal cartesian direction (i.e. 
by applying the external potential
$\delta\upsilon_{ext}({\bf r},t)=-\mathbf{k_0}\cdot\mathbf{r}\;\delta(t)$).
The magnitude of the perturbation is kept small in order to keep the dipolar
response linear\cite{oct06}. 
In this way all the frequencies of the system are excited with the same weight.
The perturbed Kohn-Sham orbitals, 
$\phi_j({\bf r},t=0^+)=\exp(i\mathbf{k_0}\cdot\mathbf{r})\tilde{\phi}_j({\bf r})$,
are then propagated on a real-time grid using the enforced
time-reversal symmetry method \cite{propagators}. 

From the Fourier transform of the real-time response to this perturbation the dynamical polarizability is obtained in the frequency range of interest, and 
finally the dipole strength function, which is proportional to the photo-absorption cross 
section, is obtained by averaging the polarizability over the three cartesian directions. More detail on this procedure can be found in Ref.~\cite{oct06}.
The time evolution of the
Kohn Sham orbitals was calculated with a time step of 1.7 {\sl attoseconds
(as)} for the triad, where a fourth order Taylor expansion approximates the
propagator. A time step of 7.9 {\sl as} (resulting in a spectral range window
for the dynamical polarizability of at least 8 {\sl eV}) could be used on the
isolated parts, where the propagator was approximated with the Lanczos method
\cite{lanczos}. The orbitals were evolved respectively for a total of 15 {\sl
fs}, that corresponds to a spectral resolution of about 0.15 {\sl eV}.

\begin{figure}[h]
\includegraphics[width=0.9\textwidth]{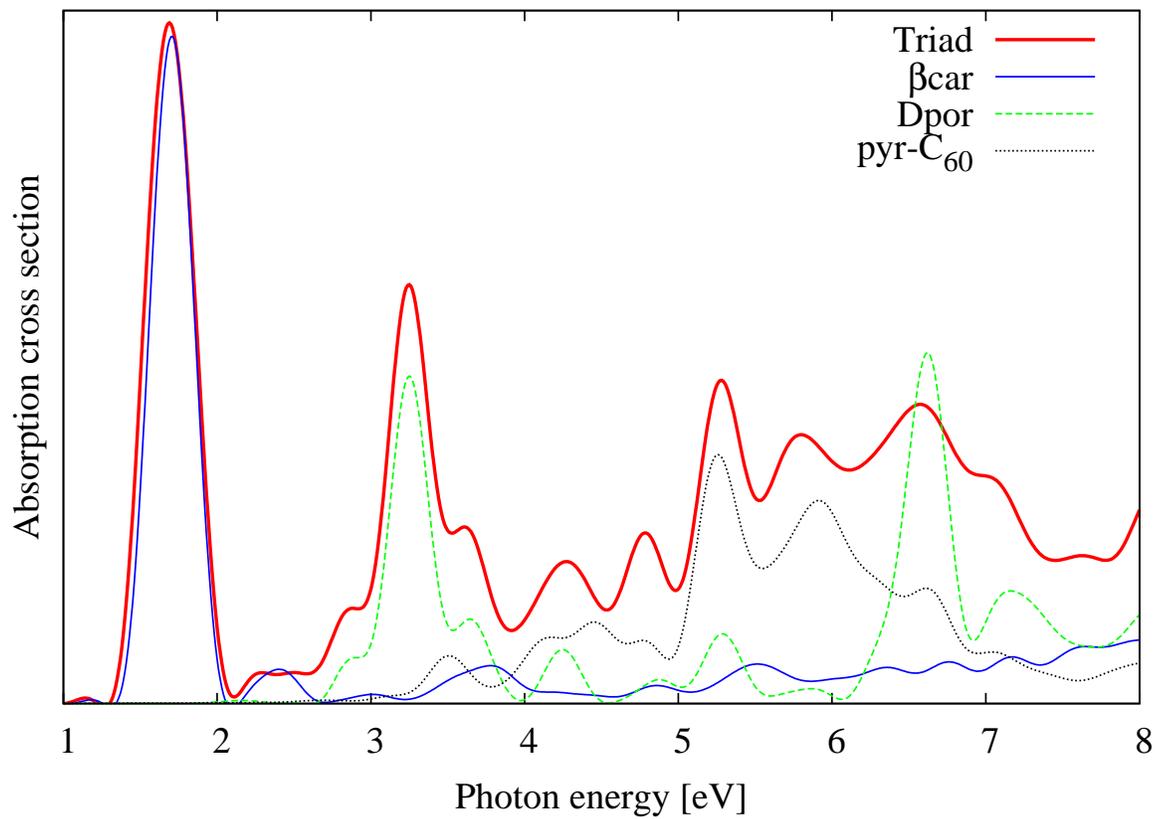}
\caption{Calculated absorption spectrum of the triad, and the isolated pyrrole-C$_{60}$, Diaryl-porphyrin and $\beta$-carotenoid.}\label{fig:abs}
\end{figure}

The TDDFT photo-absorption cross-section of the triad and of the separated
molecules is shown in~\ref{fig:abs}. We can easily distinguish one dominant peak at 1.7 {\sl eV}, several small features between 2.0 and 3.0 {\sl eV}, another peak at 3.3 with a shoulder at 3.6 {\sl eV}, and a group of peaks between 4.0 and 8.0 {\sl eV}.

The decomposition of the spectrum into optical densities of the isolated
moieties clearly shows again that the total spectrum is very well approximated
by the sum of the spectra of the parts. In particular the main feature at 1.7
{\sl eV} appears to be entirely originated by the $\beta$-carotenoid, which has
a weak contribution on the rest of the spectral range, apart from a small
feature at 2.3 {\sl eV} that partially obscures the weak Soret Q-band of the
diaryl-porphyrin. The B-band of the diaryl-porphyrin is the main contribution at
3.2 {\sl eV}, and the diaryl-porphyrin has another prominent peak in the
ultraviolet region at 6.6 {\sl eV}. The contribution of pyrrole-C$_{60}$ counts
many peaks, notably two isolated at 5.3 and 5.9 {\sl eV}, and several more from
4.0 to 5.0 {\sl eV}.

\begin{figure}[h]
\includegraphics[width=0.9\textwidth]{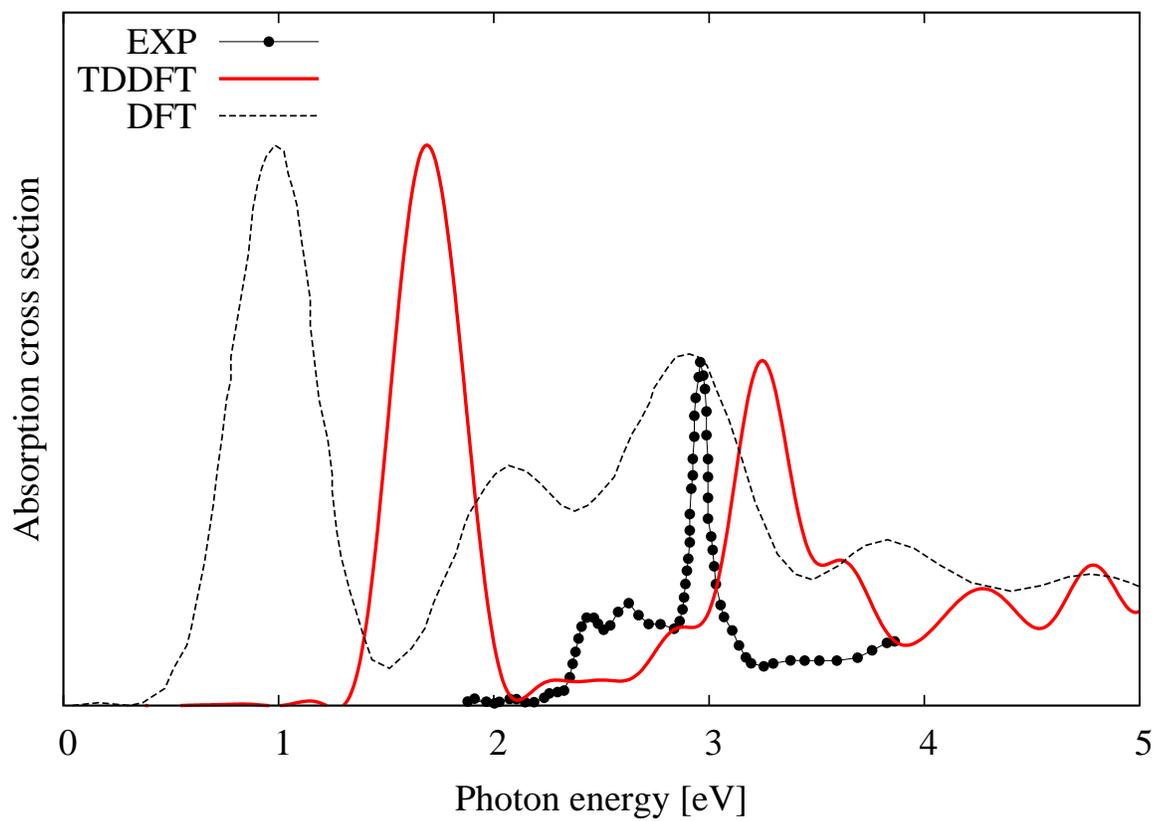}
\caption{Comparison between theoretical and observed optical absorption for the
full C-P-C$_{60}$. DFT is the calculated spectrum as in Ref.~\citep{baruah_triad}. TDDFT present work. EXP depicts the observed
spectrum as reported in Ref.~\citep{triad_synth}.}\label{fig:theo_exp}
\end{figure}

The comparison of our calculations with the experimentally observed absorption
in similar triads clearly shows that TDDFT improves on DFT in describing
the main features of the spectra. In~\ref{fig:theo_exp} we compare our data to
the absorption reported in Ref.~\citep{triad_synth}, that refers to a triad in
2-methyltetrahydrofuran solution. We observe a good one-to-one
correspondence between the calculated and the observed peaks, whose positions
appear overestimated in the calculation by an overall rigid shift of about 0.3
{\sl eV}. We do not investigate here the possible role
of the environment, that might well be responsible of the rigid shift, but is
unlikely to qualitatively change the shape of the spectrum. In
fact triads made of slightly different components, or immersed in different solvents, often have very similar spectra (see
Ref.~\citep{triads_exp}). In particular the prominent porphyrin peak around 3
{\sl eV} and the wide band between 2.0 and 3.0 {\sl eV} seem to be common
features in all the experimental data. Also notice that, despite the apparent superposition of the
main experimental peak to a peak in the DFT calculation, the nature of the
latter is predicted to be of C$_{60}$ character, while the corresponding TDDFT
feature at 3.5 correctly is attributed to the porphyrin. (see decomposition
in~\ref{fig:abs}). The experimental data range in
Refs.~\cite{triad_synth,triads_exp} are limited above approximately 2.0 {\sl eV}, but
the $\beta$-carotenoid absorption does not appear to be relevant for the
photo-conversion in the triad. On the other hand, Ref.~\cite{bcar_exp} report a
peak for the isolated $\beta$-carotene at approximately 1.24 {\em eV}. A TDDFT calculation on the same system results again in a blue shift of about 0.4 {\sl eV}. For an in depth discussion about the effect of exciton confinement in molecular chains see also Refs.~\citep{tddft_chains,todo}.

We have thus obtained the first {\sl ab initio} optical absorption of the triad
in agreement with the observed one. The comparison with a DFT spectrum, obtained
by applying the Fermi golden rule to the Kohn-Sham levels confirms (see~\ref{fig:theo_exp}) that TDDFT achieves a major improvement over DFT.

\begin{figure}[h]
\includegraphics[width=0.9\textwidth]{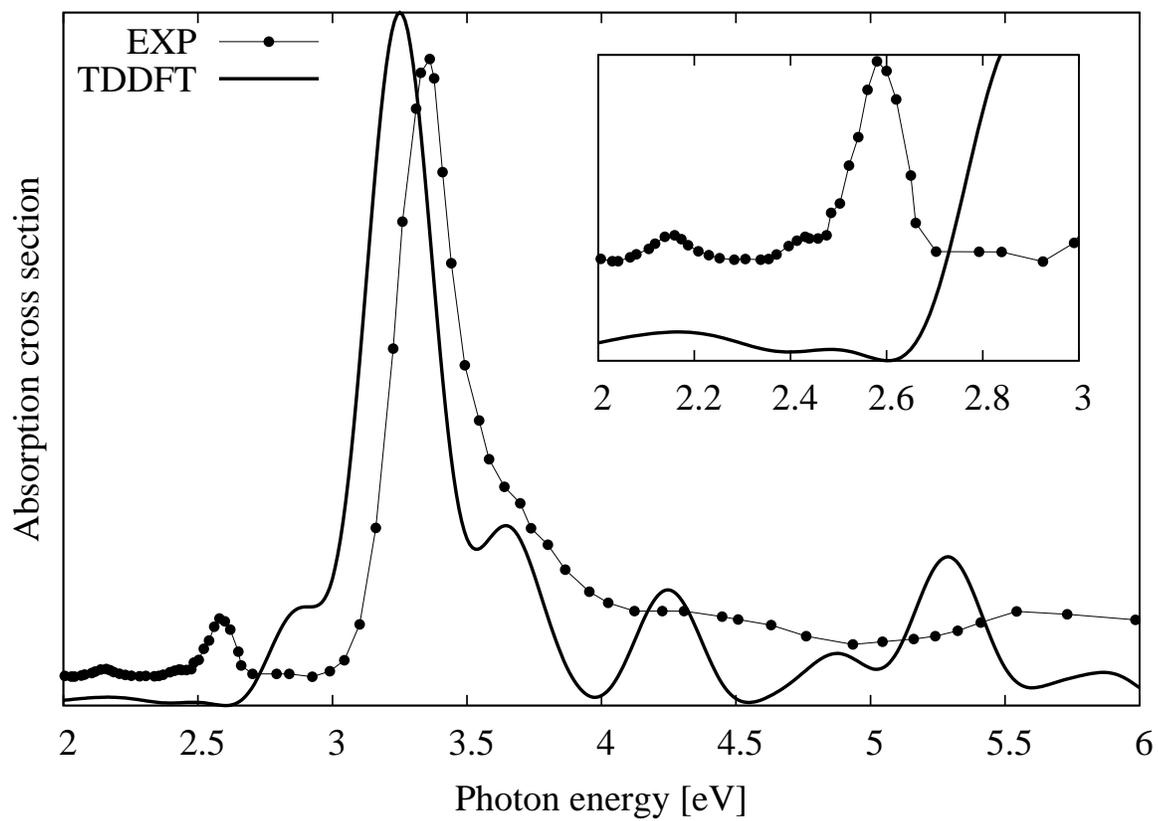}
\caption{Comparison between theoretical TDDFT optical absorption in
diaryl-porphyrin and observed free-base porphin absorption as in
Ref.~\citep{exp_porph}. In the inset the region between 2 to 3 eV is
magnified.}\label{fig:theo_exp_porph}
\end{figure}

The analysis of the absorption of the diaryl-porphyrin alone allows us to
clarify that, even if an intense peak appears in the UV region around
3.2 {\sl eV} (see~\ref{fig:theo_exp_porph}), the actual very weak onset of
absorption is in the orange-yellow visible region at approximately 2.0 {\sl
eV}. This transition corresponds to the calculated HOMO-LUMO in the
diaryl-porphyrin alone, and it is also responsible for the optical excitation
of the triad as a whole. This excitation was in fact obtained in laboratory
condition with a laser at 590 nm \cite{triad_synth}.

In~\ref{fig:triad_levels} we have emphasized, among others, the two orbitals,
localized on the porphyrin moiety, that are responsible for the optical
excitation.


On the basis of structure, level, and orbital localization we are now able to
achieve a theoretically motivated view of the photo-excitation process, and of
the subsequent charge-transfer that can be summarized in the following chain of steps:

\begin{gather*}
\textrm{C-P-C}_{60} + h\nu \to \textrm{C-P}^*\textrm{-C}^*_{60} \\
\textrm{C-P}^*\textrm{-C}^*_{60} \to \textrm{C-P}^+\textrm{-C}^{*-}_{60} \to \textrm{C}^+\textrm{-P-C}_{60}^{-}\\
\end{gather*}

The first step corresponds to the optical transition from HOMO-2 to
LUMO+4. A Casida analysis~\cite{casida1} of this transition reveals that it is
a pure transition, and the inspection of the involved orbitals clarifies that
HOMO-2 are both mostly localized on the porphyrin. Let us call this excited
state C-P$^*$-C$^*_{60}$. The electron in a previously unoccupied orbital on
the porphyrin is now under the influence of the high affinity of the C$_{60}$,
and it is easily accommodated on the almost-degenerate orbitals LUMO+1, LUMO+2
at 0.14 {\sl eV} above LUMO localized on the C$_{60}$. In the assembled triad
the three-fold degeneracy of the C$_{60}$ T$_{1u}$ LUMO is broken due to the
presence of the Pyrrole and the LUMO of the triad is a singlet state. We have
then reached a configuration C-P$^+$-C$^{*-}_{60}$. The last step consists in a
hole delocalization from the Porphyrin HOMO to the Carotenoid HOMO to give the
final state C$^+$-P-C$_{60}^{-}$. Obviously only the study of the full
dynamics of the whole process can support the latter two steps interpretation,
which is inferred here on the basis of the analysis of the energy levels in \ref{fig:triad_levels}, and on the electrochemical properties of the
moieties. The analysis of the full dynamics is quite demanding, since the processes are estimated
to occur on a time scale of tens of {\em ns}, and it is beyond the scope of the
present work.

As we have already stressed, it is the almost non-interacting character of the
moieties, demonstrated by the shape of the absorption spectrum, and the nature
of the molecular orbitals, that makes it possible to describe this process in
terms of orbitals localized on each component. In fact the charge-transfer
states associated with this molecule are not dipole allowed transitions due to
the fact that the hole on the (donor) carotene and the particle-state on the
$C_{60}$ (acceptor) molecule are essentially non overlapping. Additional work
is required to fully understand the dynamics of the light-induced charge
transfer process \cite{todo}. As concluding remarks we must recall that
the 15 {\em fs} time span of our simulation is unlikely to account for the
complete charge transfer process, that occurs on a different time scale, but it
is able to accurately describe the photo-excitation of the
molecule. Nevertheless we have been able to get a valuable insight about the
charge transfer from the careful examination of the energy levels and of the
calculated absorption spectrum of the system. Finally many issues may be relevant for a full understanding of the mechanism we have devised in the steps indicated above, and many of them still are under investigation~\cite{todo}. Among them we stress the need of a full dynamical simulation of the molecule in an intense laser field, the study of the relaxation in the excited state, and the careful study of the predictive power of the different functionals in the correct description of the charge transfer. For more information about the latter point, and the role of the self-interaction problem see also Refs.\cite{tddft_chains,pemmaraju08,korzdorfer08,ruzsinszky08}

\section{Conclusions}

In summary, we have addressed the fundamental problem of describing the creation
of electron-hole pairs and the efficient charge separation in a supra-molecular
assembly of interest for third generation solar devices. Since not all the
effects leading to the final-state charge separation are fully known the
accurate description of the excited states of such an object is a crucial step
in order to understand the details of the photo-absorption and charge-transfer
processes occurring in the system.

We have calculated for the first time the optical absorption cross section of
such a large system using TDDFT as a rigorous ground for the excited states
dynamics. In addition to the computational challenge, our calculations
demonstrate that the simple ground-state DFT description of the system is not
able to capture the correct shape of the photo-absorption spectrum. In contrast
we have shown that the main features of the TDDFT spectrum are in good
agreement with the experimental data, and the analysis of the total absorption
in terms of the absorption of the isolated moieties indicates that even at the
TDDFT level (at least in the weak field limit) the component molecules in the
triad do not appear to strongly interact.

The system as a whole acts as the sum of its parts, each of which has a specific
role in the photo-induced charge separation mechanism that makes this
molecule a good candidate for third generation solar energy conversion devices. This fact has allowed us to sketch a diagram of all the states involved in the photo-absorption and charge transfer processes, and indicates that the design of solar devices based on similar assemblies can be performed by individually addressing the active components, and then chemically joining them into a unit capable of transforming the energy of an incoming photon into electric potential energy stored into a 50 \AA\ long dipole.

%
%
%
\begin{acknowledgments}

The computer resources for the TDDFT calculations were provided by CINECA,
Bologna. TB and MRP were supported in part by ONR and NSF. Geometry
optimizations were performed on NRL and ARL computers through the support of the
DoD HPCMO. A.R. acknowledges funding by the Spanish MEC (FIS2007-65702-C02-01),
"Grupos Consolidados UPV/EHU del Gobierno Vasco" (IT-319-07), and the European
Community through NoE Nanoquanta (NMP4-CT-2004-500198), e-I3 ETSF project
(INFRA-2007-1.2.2: Grant Agreement Number 211956) and SANES
(NMP4-CT-2006-017310).

%
%
\end{acknowledgments}


\bibliography{triad-paper}

\begin{thebibliography}{25}
\expandafter\ifx\csname natexlab\endcsname\relax\def\natexlab#1{#1}\fi
\expandafter\ifx\csname bibnamefont\endcsname\relax
  \def\bibnamefont#1{#1}\fi
\expandafter\ifx\csname bibfnamefont\endcsname\relax
  \def\bibfnamefont#1{#1}\fi
\expandafter\ifx\csname citenamefont\endcsname\relax
  \def\citenamefont#1{#1}\fi
\expandafter\ifx\csname url\endcsname\relax
  \def\url#1{\texttt{#1}}\fi
\expandafter\ifx\csname urlprefix\endcsname\relax\def\urlprefix{URL }\fi
\providecommand{\bibinfo}[2]{#2}
\providecommand{\eprint}[2][]{\url{#2}}

\bibitem[{\citenamefont{Green}(2004)}]{green04}
\bibinfo{author}{\bibfnamefont{M.}~\bibnamefont{Green}},
  \emph{\bibinfo{title}{Third Generation Photovoltaics: advanced solar energy
  conversion}} (\bibinfo{publisher}{Springer-Verlag},
  \bibinfo{address}{Berlin}, \bibinfo{year}{2004}).

\bibitem[{\citenamefont{Kamat}(2007)}]{energy_review}
\bibinfo{author}{\bibfnamefont{P.~V.} \bibnamefont{Kamat}},
  \bibinfo{journal}{J. Phys. Chem. C} \textbf{\bibinfo{volume}{111}},
  \bibinfo{pages}{2834} (\bibinfo{year}{2007}).

\bibitem[{\citenamefont{Gust et~al.}(2001)\citenamefont{Gust, Moore, and
  Moore}}]{photosynthesis}
\bibinfo{author}{\bibfnamefont{D.}~\bibnamefont{Gust}},
  \bibinfo{author}{\bibfnamefont{T.~A.} \bibnamefont{Moore}}, \bibnamefont{and}
  \bibinfo{author}{\bibfnamefont{A.~L.} \bibnamefont{Moore}},
  \bibinfo{journal}{Acc. Chem. Res.} \textbf{\bibinfo{volume}{34}},
  \bibinfo{pages}{40} (\bibinfo{year}{2001}).

\bibitem[{\citenamefont{O'Regan and Gr\"atzel}(1991)}]{gratzel91}
\bibinfo{author}{\bibfnamefont{B.}~\bibnamefont{O'Regan}} \bibnamefont{and}
  \bibinfo{author}{\bibfnamefont{M.}~\bibnamefont{Gr\"atzel}},
  \bibinfo{journal}{Nature} \textbf{\bibinfo{volume}{335}},
  \bibinfo{pages}{737} (\bibinfo{year}{1991}).

\bibitem[{\citenamefont{Gr\"atzel}(2001)}]{gratzel01}
\bibinfo{author}{\bibfnamefont{M.}~\bibnamefont{Gr\"atzel}},
  \bibinfo{journal}{Nature} \textbf{\bibinfo{volume}{414}},
  \bibinfo{pages}{338} (\bibinfo{year}{2001}).

\bibitem[{\citenamefont{Guldi}(2002)}]{antenna}
\bibinfo{author}{\bibfnamefont{D.~M.} \bibnamefont{Guldi}},
  \bibinfo{journal}{Chem. Soc. Rev.} \textbf{\bibinfo{volume}{31}},
  \bibinfo{pages}{22} (\bibinfo{year}{2002}).

\bibitem[{\citenamefont{Baruah and Pederson}(2006)}]{baruah_triad}
\bibinfo{author}{\bibfnamefont{T.}~\bibnamefont{Baruah}} \bibnamefont{and}
  \bibinfo{author}{\bibfnamefont{M.~R.} \bibnamefont{Pederson}},
  \bibinfo{journal}{J. Chem. Phys.} \textbf{\bibinfo{volume}{125}},
  \bibinfo{pages}{164706} (\bibinfo{year}{2006}).

\bibitem[{\citenamefont{Runge and Gross}(1984)}]{tddft}
\bibinfo{author}{\bibfnamefont{E.}~\bibnamefont{Runge}} \bibnamefont{and}
  \bibinfo{author}{\bibfnamefont{E.~K.~U.} \bibnamefont{Gross}},
  \bibinfo{journal}{Phys. Rev. Lett.} \textbf{\bibinfo{volume}{52}},
  \bibinfo{pages}{997} (\bibinfo{year}{1984}).

\bibitem[{\citenamefont{Perdew and Zunger}(1981)}]{perdew-zunger81}
\bibinfo{author}{\bibfnamefont{P.}~\bibnamefont{Perdew}} \bibnamefont{and}
  \bibinfo{author}{\bibfnamefont{A.}~\bibnamefont{Zunger}},
  \bibinfo{journal}{Phys. Rev. B} \textbf{\bibinfo{volume}{23}},
  \bibinfo{pages}{5048} (\bibinfo{year}{1981}).

\bibitem[{\citenamefont{Troullier and Martins}(1991)}]{troullier-martins91}
\bibinfo{author}{\bibfnamefont{N.}~\bibnamefont{Troullier}} \bibnamefont{and}
  \bibinfo{author}{\bibfnamefont{J.~L.} \bibnamefont{Martins}},
  \bibinfo{journal}{Phys. Rev. B} \textbf{\bibinfo{volume}{43}},
  \bibinfo{pages}{8861} (\bibinfo{year}{1991}).

\bibitem[{\citenamefont{Genovese et~al.}(2006)\citenamefont{Genovese, Deutsch,
  Neelov, Goedecker, and Beylkin}}]{isf}
\bibinfo{author}{\bibfnamefont{L.}~\bibnamefont{Genovese}},
  \bibinfo{author}{\bibfnamefont{T.}~\bibnamefont{Deutsch}},
  \bibinfo{author}{\bibfnamefont{A.}~\bibnamefont{Neelov}},
  \bibinfo{author}{\bibfnamefont{S.}~\bibnamefont{Goedecker}},
  \bibnamefont{and} \bibinfo{author}{\bibfnamefont{G.}~\bibnamefont{Beylkin}},
  \bibinfo{journal}{J. Chem. Phys.} \textbf{\bibinfo{volume}{125}},
  \bibinfo{pages}{74105} (\bibinfo{year}{2006}).

\bibitem[{\citenamefont{Castro et~al.}(2006)\citenamefont{Castro, Appel,
  Oliveira, Rozzi, Andrade, Lorenzen, Marques, Gross, and Rubio}}]{oct06}
\bibinfo{author}{\bibfnamefont{A.}~\bibnamefont{Castro}},
  \bibinfo{author}{\bibfnamefont{H.}~\bibnamefont{Appel}},
  \bibinfo{author}{\bibfnamefont{M.}~\bibnamefont{Oliveira}},
  \bibinfo{author}{\bibfnamefont{C.~A.} \bibnamefont{Rozzi}},
  \bibinfo{author}{\bibfnamefont{X.}~\bibnamefont{Andrade}},
  \bibinfo{author}{\bibfnamefont{F.}~\bibnamefont{Lorenzen}},
  \bibinfo{author}{\bibfnamefont{M.~A.~L.} \bibnamefont{Marques}},
  \bibinfo{author}{\bibfnamefont{E.~K.~U.} \bibnamefont{Gross}},
  \bibnamefont{and} \bibinfo{author}{\bibfnamefont{A.}~\bibnamefont{Rubio}},
  \bibinfo{journal}{Phys. Stat. Sol. B} \textbf{\bibinfo{volume}{243}},
  \bibinfo{pages}{2465} (\bibinfo{year}{2006}).

\bibitem[{\citenamefont{Marques et~al.}(2006)\citenamefont{Marques, Ullrich,
  Nogueira, Rubio, Burke, and Gross}}]{tddftbook}
\bibinfo{editor}{\bibfnamefont{M.~A.~L.} \bibnamefont{Marques}},
  \bibinfo{editor}{\bibfnamefont{C.}~\bibnamefont{Ullrich}},
  \bibinfo{editor}{\bibfnamefont{F.}~\bibnamefont{Nogueira}},
  \bibinfo{editor}{\bibfnamefont{A.}~\bibnamefont{Rubio}},
  \bibinfo{editor}{\bibfnamefont{K.}~\bibnamefont{Burke}}, \bibnamefont{and}
  \bibinfo{editor}{\bibfnamefont{E.~K.~U.} \bibnamefont{Gross}}, eds.,
  \emph{\bibinfo{title}{Time-dependent density functional theory}}
  (\bibinfo{publisher}{Springer-Verlag}, \bibinfo{address}{Berlin},
  \bibinfo{year}{2006}).

\bibitem[{\citenamefont{Castro et~al.}(2004)\citenamefont{Castro, Marques, and
  Rubio}}]{propagators}
\bibinfo{author}{\bibfnamefont{A.}~\bibnamefont{Castro}},
  \bibinfo{author}{\bibfnamefont{M.~A.~L.} \bibnamefont{Marques}},
  \bibnamefont{and} \bibinfo{author}{\bibfnamefont{A.}~\bibnamefont{Rubio}},
  \bibinfo{journal}{J. Chem. Phys} \textbf{\bibinfo{volume}{121}},
  \bibinfo{pages}{3425} (\bibinfo{year}{2004}).

\bibitem[{\citenamefont{Hochbruck and Lubich}(1997)}]{lanczos}
\bibinfo{author}{\bibfnamefont{M.}~\bibnamefont{Hochbruck}} \bibnamefont{and}
  \bibinfo{author}{\bibfnamefont{C.}~\bibnamefont{Lubich}},
  \bibinfo{journal}{SIAM J. Numer. Anal.} \textbf{\bibinfo{volume}{34}},
  \bibinfo{pages}{1911} (\bibinfo{year}{1997}).

\bibitem[{\citenamefont{Kodis et~al.}(2004)\citenamefont{Kodis, Liddell, Moore,
  Moore, and Gust}}]{triad_synth}
\bibinfo{author}{\bibfnamefont{G.}~\bibnamefont{Kodis}},
  \bibinfo{author}{\bibfnamefont{P.~A.} \bibnamefont{Liddell}},
  \bibinfo{author}{\bibfnamefont{A.~L.} \bibnamefont{Moore}},
  \bibinfo{author}{\bibfnamefont{T.~A.} \bibnamefont{Moore}}, \bibnamefont{and}
  \bibinfo{author}{\bibfnamefont{D.}~\bibnamefont{Gust}}, \bibinfo{journal}{J.
  Phys. Org. Chem.} \textbf{\bibinfo{volume}{17}}, \bibinfo{pages}{724}
  (\bibinfo{year}{2004}).

\bibitem[{\citenamefont{Rizzi et~al.}(2008)\citenamefont{Rizzi, van Gastel,
  Liddell, Palacios, Moore, Kodis, Moore, Moore, Gust, and
  Braslavsky}}]{triads_exp}
\bibinfo{author}{\bibfnamefont{A.~C.} \bibnamefont{Rizzi}},
  \bibinfo{author}{\bibfnamefont{M.}~\bibnamefont{van Gastel}},
  \bibinfo{author}{\bibfnamefont{P.~A.} \bibnamefont{Liddell}},
  \bibinfo{author}{\bibfnamefont{R.~E.} \bibnamefont{Palacios}},
  \bibinfo{author}{\bibfnamefont{G.~F.} \bibnamefont{Moore}},
  \bibinfo{author}{\bibfnamefont{G.}~\bibnamefont{Kodis}},
  \bibinfo{author}{\bibfnamefont{A.~L.} \bibnamefont{Moore}},
  \bibinfo{author}{\bibfnamefont{T.~A.} \bibnamefont{Moore}},
  \bibinfo{author}{\bibfnamefont{D.}~\bibnamefont{Gust}}, \bibnamefont{and}
  \bibinfo{author}{\bibfnamefont{S.~E.} \bibnamefont{Braslavsky}},
  \bibinfo{journal}{J. Phys. Chem. A} \textbf{\bibinfo{volume}{112}},
  \bibinfo{pages}{421} (\bibinfo{year}{2008}).

\bibitem[{\citenamefont{Jeevarajan et~al.}(1994)\citenamefont{Jeevarajan,
  Kispert, and Wu}}]{bcar_exp}
\bibinfo{author}{\bibfnamefont{A.~S.} \bibnamefont{Jeevarajan}},
  \bibinfo{author}{\bibfnamefont{L.~D.} \bibnamefont{Kispert}},
  \bibnamefont{and} \bibinfo{author}{\bibfnamefont{X.}~\bibnamefont{Wu}},
  \bibinfo{journal}{Chem. Phys. Lett.} \textbf{\bibinfo{volume}{219}},
  \bibinfo{pages}{427} (\bibinfo{year}{1994}).

\bibitem[{\citenamefont{Varsano et~al.}(2008)\citenamefont{Varsano, Marini, and
  Rubio}}]{tddft_chains}
\bibinfo{author}{\bibfnamefont{D.}~\bibnamefont{Varsano}},
  \bibinfo{author}{\bibfnamefont{A.}~\bibnamefont{Marini}}, \bibnamefont{and}
  \bibinfo{author}{\bibfnamefont{A.}~\bibnamefont{Rubio}},
  \bibinfo{journal}{Phys. Rev. Lett.} \textbf{\bibinfo{volume}{101}},
  \bibinfo{pages}{133002} (\bibinfo{year}{2008}).

\bibitem[{\citenamefont{Spallanzani et~al.}()\citenamefont{Spallanzani, Rozzi,
  Manghi, and Rubio}}]{todo}
\bibinfo{author}{\bibfnamefont{N.}~\bibnamefont{Spallanzani}},
  \bibinfo{author}{\bibfnamefont{C.~A.} \bibnamefont{Rozzi}},
  \bibinfo{author}{\bibfnamefont{F.}~\bibnamefont{Manghi}}, \bibnamefont{and}
  \bibinfo{author}{\bibfnamefont{A.}~\bibnamefont{Rubio}}, \bibinfo{note}{in
  preparation}.

\bibitem[{\citenamefont{Casida}(1995)}]{casida1}
\bibinfo{author}{\bibfnamefont{M.~E.} \bibnamefont{Casida}}, in
  \emph{\bibinfo{booktitle}{Recent Advances in Density Functional Methods}},
  edited by \bibinfo{editor}{\bibfnamefont{D.~P.} \bibnamefont{Chong}}
  (\bibinfo{publisher}{World Scientific Press}, \bibinfo{address}{Singapore},
  \bibinfo{year}{1995}), vol.~\bibinfo{volume}{I}, p. \bibinfo{pages}{155}.

\bibitem[{\citenamefont{Pemmaraju et~al.}(2008)\citenamefont{Pemmaraju,
  Sanvito, and Burke}}]{pemmaraju08}
\bibinfo{author}{\bibfnamefont{C.~D.} \bibnamefont{Pemmaraju}},
  \bibinfo{author}{\bibfnamefont{S.}~\bibnamefont{Sanvito}}, \bibnamefont{and}
  \bibinfo{author}{\bibfnamefont{K.}~\bibnamefont{Burke}},
  \bibinfo{journal}{Phys. Rev. B} \textbf{\bibinfo{volume}{77}},
  \bibinfo{pages}{121204} (\bibinfo{year}{2008}).

\bibitem[{\citenamefont{K{\"o}rzd{\"o}rfer
  et~al.}(2008)\citenamefont{K{\"o}rzd{\"o}rfer, Mundt, and
  K{\"u}mmel}}]{korzdorfer08}
\bibinfo{author}{\bibfnamefont{T.}~\bibnamefont{K{\"o}rzd{\"o}rfer}},
  \bibinfo{author}{\bibfnamefont{M.}~\bibnamefont{Mundt}}, \bibnamefont{and}
  \bibinfo{author}{\bibfnamefont{S.}~\bibnamefont{K{\"u}mmel}},
  \bibinfo{journal}{Phys. Rev. Lett.} \textbf{\bibinfo{volume}{100}},
  \bibinfo{pages}{133004} (\bibinfo{year}{2008}).

\bibitem[{\citenamefont{Ruzsinszky et~al.}(2008)\citenamefont{Ruzsinszky,
  Perdew, Csonka, Scuseria, and Vydrov}}]{ruzsinszky08}
\bibinfo{author}{\bibfnamefont{A.}~\bibnamefont{Ruzsinszky}},
  \bibinfo{author}{\bibfnamefont{J.~P.} \bibnamefont{Perdew}},
  \bibinfo{author}{\bibfnamefont{G.~I.} \bibnamefont{Csonka}},
  \bibinfo{author}{\bibfnamefont{G.~E.} \bibnamefont{Scuseria}},
  \bibnamefont{and} \bibinfo{author}{\bibfnamefont{O.~A.}
  \bibnamefont{Vydrov}}, \bibinfo{journal}{Phys. Rev. A}
  \textbf{\bibinfo{volume}{77}}, \bibinfo{pages}{060502}
  (\bibinfo{year}{2008}).

\bibitem[{\citenamefont{Edwards et~al.}(1971)\citenamefont{Edwards, Dolphin,
  Gouterman, and Adler}}]{exp_porph}
\bibinfo{author}{\bibfnamefont{L.}~\bibnamefont{Edwards}},
  \bibinfo{author}{\bibfnamefont{D.~H.} \bibnamefont{Dolphin}},
  \bibinfo{author}{\bibfnamefont{M.}~\bibnamefont{Gouterman}},
  \bibnamefont{and} \bibinfo{author}{\bibfnamefont{A.~D.} \bibnamefont{Adler}},
  \bibinfo{journal}{J. Mol. Spect.} \textbf{\bibinfo{volume}{38}},
  \bibinfo{pages}{16} (\bibinfo{year}{1971}).

\end{thebibliography}


\begin{table}[h]
\begin{tabular}{ccc}
\multicolumn{3}{c}{Comparing HOMO and LUMO energies}\\
\multicolumn{3}{c}{}\\\hline
molecules & $\epsilon_{HOMO}$ (eV) & $\epsilon_{LUMO}$ (eV) \\ \hline\hline
$\beta$-carotenoid & -4.47 & -3.49 \\
Diaryl-porphyrin   & -5.15 & -3.31 \\
pyrrole-C$_{60}$   & -5.89 & -4.42 \\
Triad              & -4.51 & -4.38 \\ \hline
\end{tabular}
\caption{Kohn-Sham energy of HOMO and LUMO respectively for $\beta$-carotenoid,
Diaryl-porphyrin, pyrrole-C$_{60}$ and the whole triad.}\label{tab:en_lev}
\end{table}


\newpage

\newpage

\newpage

\newpage



\newpage
\begin{figure}[!t]
\vspace{7cm}
\includegraphics{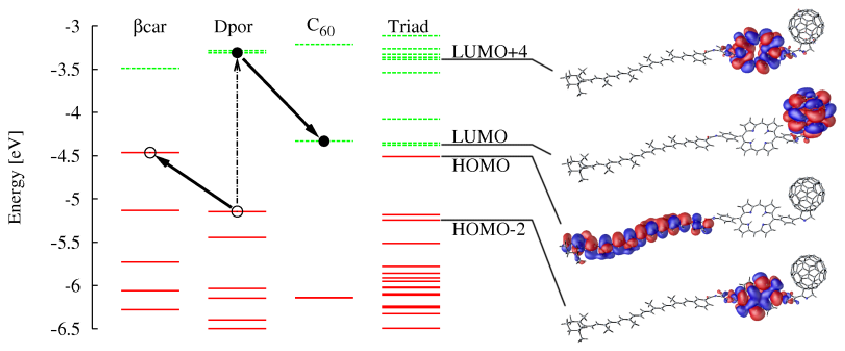}
\end{figure}

\end{document}